# Cardiac and respiratory motion extraction for MRI using Pilot Tone–a patient study

**Authors Names and Degrees:**

Chong Chen[1] PhD, Yingmin Liu[2] PhD, Orlando P. Simonetti[2] PhD, Matthew Tong[2] MD, Ning Jin[3] PhD, Mario Bacher[4] MS, Peter Speier[4] PhD and Rizwan Ahmad[1] PhD

**Author Affiliations:**

1. Department of Biomedical Engineering, The Ohio State University, Columbus, US.; 2. Davis Heart & Lung Research Institute, The Ohio State University, Columbus, US.; 3. Siemens Medical Solutions USA, Inc., Columbus, US.; 4. Siemens Healthcare GmbH, Erlangen, Germany, Erlangen, Germany.

**Acknowledgements:** Not applicable

**Grant Support:** National Institutes of Health R01HL135489 and R01HL151697

**Running Title:** Cardiac and respiratory motion extraction for MRI using Pilot Tone–a patient study

# Cardiac and respiratory motion extraction for MRI using Pilot Tone–a patient study

## Abstract

**Background**: Several studies have shown that both respiratory and cardiac motion can be extracted from the Pilot Tone (PT) signal successfully. However, most of these studies were performed in healthy volunteers. In addition, validating PT using ECG as a reference can be problematic because both PT and ECG tend to be unreliable in patients with arrhythmias.

**Purpose:** We seek to evaluate the accuracy and reliability of the cardiac and respiratory signals extracted from PT in patients clinically referred for cardiovascular MRI with the image-derived signals as the reference.

**Study type:** Retrospective

**Subjects:** Twenty-three patients

**Field strength/sequence**: 1.5 T scanner, balanced steady-state free-precession real-time (RT) cine sequence. The PT signal was generated by a built-in PT transmitter integrated within the body array coil. For comparison, commercial ECG and BioMatrix (BM) respiratory sensor signals were synchronously recorded.

**Statistical Test:** Pearson correlation, mean absolute difference, linear regression.

**Results**: The respiratory motion extracted from PT correlated positively with the image-derived respiratory signal in all cases and showed a stronger correlation (absolute coefficient: 0.95 ± 0.09) than BM (0.72 ± 0.24). For the cardiac signal, PT trigger jitter (standard deviation of PT trigger locations relative to ECG triggers) ranged from 6.6 to 83.3 ms, with a median of 21.8 ms. The mean absolute difference between the PT and corresponding ECG cardiac cycle duration was less than 5% of the averaged ECG RR interval for 21 out of 23 patients. Overall, the performance of PT-based trigger extraction was comparable to that of ECG. We did not observe significant linear dependence ($p > 0.28$) of PT delay and PT jitter on the patients' BMI or cardiac cycle duration.

**Conclusions**: This study demonstrates the potential of PT to monitor both respiratory and cardiac motion in patients clinically referred for cardiovascular MRI.

## Introduction

Cardiovascular magnetic resonance imaging (CMR) is a non-invasive technique that provides a comprehensive assessment of the cardiovascular system and is widely used as a diagnostic tool. A single CMR exam can assess the structure, function, and morphology of the heart as well as quantitatively evaluate hemodynamics in the valves and vessels. A major challenge in the clinical utility of CMR is its inherently inefficient acquisition, which not only prolongs the exam time but also makes the image quality sensitive to both respiratory and cardiac motion. Therefore, monitoring respiratory and cardiac motion is considered an integral part of CMR, and almost all CMR techniques take physiological motion into account at the acquisition or reconstruction stage.

A simple way of minimizing artifacts from respiratory motion is to suspend respiration during the scan. However, breath-holding can be challenging for patients with poor respiratory control and shortness of breath. Alternatively, navigator echoes can be employed for prospective respiratory gating [1], where the observed diaphragm position serves as a surrogate for the respiratory state and limits acquisition to a specific respiratory phase; however, setting up navigator echoes can be time-consuming. Also, since the navigator echo acquisition disrupts the steady-state of magnetization, it is not compatible with steady-state free-precession (SSFP) sequences, which are commonly used for CMR. Respiratory motion can also be captured using respiratory bellows [2]. However, this method needs prior setup and has not proven reliable for routine clinical use. In the last two decades, several self-gating (SG) techniques have been developed where the respiratory signal can be extracted post-acquisition from the MRI data and used for retrospective binning [3]. Nevertheless, SG requires modifying the pulse sequence and may lower the acquisition efficiency to accommodate periodic sampling of the k-space center or increased repetition time [4].

To monitor cardiac activity, the electrocardiogram (ECG) is commonly used in CMR. In a typical setup, the data acquisition is synchronized with the ECG R-wave to either freeze the cardiac motion (e.g., in single-shot imaging) by limiting the acquisition to a narrow temporal window in each heartbeat or deliver motion-resolved imaging (e.g., in cine). However, setting up ECG adds to the patient preparation time, and the ECG signal is prone to artifacts from rapid magnetic gradient switching and the magnetohydrodynamic effect, especially at higher field strength [5]. Additionally, the concordance between the electrical activity provided by ECG and the mechanical motion of the heart can be degraded in certain cardiac diseases [6], making the ECG a poor surrogate for the mechanical motion imaged by CMR. With additional refinements in data acquisition and processing, more recently proposed SG methods are capable of extracting the cardiac signal in addition to the respiratory signal [7,8].

Pilot Tone (PT), a contactless technique to monitor physiological motions for CMR, has been proposed recently [9,10]. The PT signal is generated by a separate radio frequency transmitter placed in close proximity to the body and is modulated by physiological motion. The frequency of the PT signal is adjusted so that it is outside the frequency range of the image but still within the receiver coil bandwidth. Therefore, the PT and MRI signals are both captured simultaneously by the receiver coils. Due to the frequency difference between the PT and MRI signals, the two signals can be separated. When integrated into the body array coil, the PT generator can eliminate the setup time associated with ECG, navigator echoes, and respiratory bellows.

It has been shown that both respiratory [11–14] and cardiac [14–17] motions can be extracted using PT. However, all these studies were performed in healthy volunteers except for [14,17]. In the work by Falcão et al. [14], PT was used for distributing 5D flow imaging data into cardiac and respiratory bins; in the work by Bacher et al. [17], PT-based prospective cardiac triggering was compared against ECG. However, the quality of PT-based respiratory and cardiac signals was not explicitly evaluated in these studies. Also, correlating PT to ECG, without an independent reference, can be problematic because both PT and ECG tend to be unreliable in patients with arrhythmias.

In this study, we extend the processing framework reported in [14,15] to extract both cardiac and respiratory signals from multi-coil PT data and then evaluate the accuracy and reliability of the extracted signals in patients referred for clinical CMR. Cardiac and respiratory signals extracted from the RT cine images were used as the ground truth.

For comparison, the commercial ECG and BioMatrix (BM) [18] respiratory sensor signals were synchronously recorded with the MRI acquisition. The BM sensor tracks the respiratory motion by monitoring respiration-induced changes in the loading of an additional coil, which is integrated into the spine coil array, placed in the vicinity of diaphragm of the patient and works at a different frequency from the Larmor frequency [18].

## Methods

### Data acquisition

Twenty-three patients (14 males, 50±15 years) referred for CMR for a variety of indications were included in this study. Patient demographics and the corresponding CMR indications are summarized in Table 1. The study was approved by the Institutional Review Board, and written informed consent was obtained from all subjects. For each patient, a short (22 out of 23 subjects) or long axis stack with 6-14 slices covering the whole heart was acquired under free-breathing conditions using a prototype balanced SSFP RT cine research sequence on a 1.5T scanner (MAGNETOM Sola, Siemens Healthcare, Erlangen, Germany). Eight to 12 coils in a 12-channel body array and 8 to 16 coils in the posterior spine array were used in the exams, resulting in 16 to 28 receiver coils in total. The other scan parameters were: TE/TR ~1.2/2.8 ms, flip angle ~ 70º, spatial resolution ~2 × 2 mm$^2$, temporal resolution ~50 ms, acquisition time 10 s/slice, and acceleration rate ~8 with pseudo-random sampling [19].

The PT signal was generated by a built-in transmitter integrated within the 12-channel body array coil. Based on the readout duration and gradient strength, the radio frequency of the PT signal was automatically adjusted for each scan to fall into the oversampling region of the acquired field-of-view (FOV). As a result, no extra time was needed to set up the PT. To facilitate a comparison, ECG and BM signals were also recorded synchronously with the MRI acquisition. The patient is positioned to ensure that the diaphragm is approximately above the BM sensor. A small current is fed to the BM sensor, which generates a reflection that encodes the respiratory information, with a minor increase during inhalation and decrease during exhalation [18]. For ECG, the R-wave triggers were determined by the scanner software. The RT cine images were reconstructed inline with a prototype Gadgetron-based [20] implementation of SCoRe algorithm [21], which is a parameter-free, SENSE-based compressed sensing reconstruction method.

**Respiratory and cardiac motion extractions**

The PT signal was separated from the MRI signal at the hardware level via quadrature detection, yielding one sample of complex-valued PT signal for every k-space line (~ 2.8 ms). The multi-coil PT signal had the dimension of $N \times C$, where $N$ is the total number of k-space lines across all acquired frames and $C$ is the number of receiver coils. The PT signal from a single receiver coil encodes an unknown linear mixture of modulations due to bulk motion, respiration, and cardiac pulsation. This linear mixture changes from coil to coil due to the relative differences in the positions of the coils with respect to the anatomy and direction of motion. Extracting cardiac and respiratory signals from the multi-coil PT data can be considered a blind source separation problem, which has been well-studied in signal processing.

For each subject, the PT signals from all the slices were concatenated along the temporal dimension before further processing. Fig. 1 shows the processing pipeline used in this study to extract respiratory and cardiac motion from the multi-coil PT signal. First, we concatenated the real and imaginary parts of the PT signal to generate an $N \times 2C$ matrix. The respiratory motion was extracted by applying [0,0.8] Hz low-pass temporal filtering followed by principal component analysis (PCA). Out of the first two principal components (PC), the component with the higher peak between [0.1,0.6] Hz was selected as the respiratory signal. The PCA-based methods, however, generate signals with arbitrary polarity. To correctly identify the expiration (E) and inspiration (I), the polarity of the recovered respiratory signal was adjusted such that it matched the polarity of the PT signal from the spine coils closest to the feet. Due to the predictable motion of abdominal organs with respect to the imaginary lines that connect the PT generator with these spine coils, the raw PT signal from these coils was observed to have a deterministic relationship with the directionality of the respiratory signal.

To extract cardiac motion, PCA was followed by fast independent component analysis (ICA) [14,15,22]. The independent component with the highest peak in the [0.5,3] Hz frequency band was automatically selected as the cardiac signal. Then, the signal was further processed using a [0.5,3] Hz band-pass filter to suppress noise, and the directionality of the cardiac signal was determined based on the assumption that the duration of diastole is longer than systole. Finally, the cardiac trigger points were detected as the negative peak in the first-order finite-difference of the selected ICA component.

**Evaluation**

To evaluate the performance of PT, we extracted the cardiac and respiratory information from the RT cine images and used it as the ground truth. The respiratory signal and its directionality were determined from RT cine images by a recently proposed PCA-based method [23]. The directionality of the signal was verified and corrected (when needed) manually using the temporal profile along the line drawn across the chest wall or diaphragm on the cine images. For each 10 s acquisition from each slice, Pearson correlation with the reference was calculated both for the PT and the built-in BM signals. In addition, the consistency of the respiratory binning between the PT signal and the image-derived reference was also calculated across all the patients.

For cardiac motion, the triggers derived from the PT ($t_{PT}$) were paired with the corresponding ECG R-wave triggers ($t_{ECG}$) using a two-step procedure. First, each PT trigger was paired with the nearest ECG trigger that was within half of the mean ECG RR interval ($RR$). Then, the mean PT trigger delay ($\bar{t}_{\text{delay}}$) relative to the ECG was estimated. Second, the PT triggers were again paired with the corresponding ECG triggers that had the smallest $|t_{PT} - t_{ECG} - \bar{t}_{\text{delay}}| \leq 0.25RR$. The first step was included to account for the delay between ECG and PT signals before identifying trigger pairs. After the pairing process, the mean and standard deviation (std) of the PT delay ($t_{PT}-t_{ECG}$) across all paired triggers were reported for each patient. For all the unpaired PT and ECG triggers, the corresponding RT cine images were blindly evaluated by a cardiologist to detect peak systolic frames. Based on this information, we categorized the unpaired triggers into five groups: false positive/negative ECG, false positive/negative PT, or indeterminate.

# Results

**Respiratory motion**

Across all slices and patients, Fig. 2 shows the correlations of the respiratory signals from PT and BM with the image-derived reference. Overall, PT-derived signal positively correlates with the reference for all slices and exhibits a stronger correlation (with absolute coefficient: 0.95±0.09) than BM (0.72±0.24). As highlighted by the red arrows, there were only six slices (out of 287) where PT poorly correlated (absolute correlation coefficient less than 0.7) with the reference. In contrast, BM exhibited a mix of positive and negative correlations with the reference, with the absolute value of the correlation coefficients less than 0.7 for 81 slices.

Fig. 3 shows exemplary respiratory signals from PT, BM, and the image-derived reference from six representative patients. The temporal profiles along the green dashed lines drawn through the diaphragm are shown for visual reference. The PT signal matches with the reference for all patients in the top two rows. BM demonstrates a high correlation with the reference for patient #2 but exhibits some delay for patient #8 and incorrect directionality for patient #10. For patient #9, the BM signal is completely distorted and has a weak correlation with the image-derived respiratory signal. The bottom row in Fig. 3 shows two examples where PT poorly correlates with the reference. There is almost no respiratory motion present for patient #21, leading to poor correlations for both PT and BM. However, both signals accurately capture the lack of respiratory motion. For patient #18, significant mismatches with the reference are observed both for PT and BM.

Table 2 shows the consistency of the respiratory binning between the PT-derived and image-derived respiratory signals for all the patients. The percentage of the readout lines where PT-based binning agrees with that from image-based binning is 80.03%, 69.05%, 75.79 and 88.85% for end-expiratory, mid-expiratory, mid-inspiratory, and end-inspiratory bins, respectively. As for the mismatched cases, most of them were off by one bin.

**Cardiac Motion**

Fig. 4 compares cardiac signals from PT and ECG for one representative patient. The temporal profile along the green dashed line drawn through the middle of the heart is shown for visual reference, with the end-diastolic frames marked with the green triangles. The noisy blue curve in the middle row is the intermediate PT signal after PCA and fast ICA. It clearly shows a periodic modulation that coincides with the cardiac motion shown in the temporal profile. The final PT signal is obtained by first filtering and then taking the first-order finite-difference of the intermediate signal. For the patient shown in Fig. 4, both PT and ECG signals have a high correlation with the cardiac motion shown in the temporal profile.

Fig. 5 demonstrates the pairing process of PT and ECG triggers. As shown in Fig. 5(a), most of the PT triggers have been successfully paired with the corresponding ECG triggers. The PT delay, defined as ($t_{PT} - t_{ECG}$), is calculated for all paired triggers. Fig. 5(b) shows that all the unpaired triggers are sorted into five groups based on the number of heartbeats recognized by a cardiologist from the corresponding cine images: false positive/negative ECG, false positive/negative PT, or indeterminate.

Table 3 summarizes the mean ECG RR interval, the total number of triggers, the number of false positive/negative PT/ECG triggers, and the mean ± standard deviation (std) of PT delay for each patient. As shown in the table, the extent of mean PT delay is patient-dependent and ranges from 79.3 to 220.1 ms. However, for a majority of subjects (16 out of 23), the delay resides in a narrow band of 150 ± 35 ms. PT trigger jitter (std of PT delay) ranges from 6.6 to 83.3 ms, with a median of 21.8 ms and most values (19 out of 23) below 33 ms. The total numbers of false ECG/PT triggers are summarized in the last row, with PT having fewer false triggers than ECG. The dependence of PT delay and PT jitter on the patients' BMI/cardiac cycle duration is shown in Supporting Information Fig. S1, where patient #21 is excluded due to high arrhythmic load and weak cardiac contraction. No significant linear relationship is observed, with the estimated linear correlation coefficient not significantly ($p > 0.28$) different from zero.

Fig. 6 demonstrates the direct comparison of the cardiac cycle duration estimated from the PT and corresponding ECG triggers. Overall, the PT cardiac cycle duration is in good agreement with ECG, with the mean absolute difference less than 5% of the averaged ECG RR interval, except in patient #21 (14.01%) and patient #22 (7.7%). For the patients with large ECG RR interval variation (#2, #6, #7, #10, #20), the PT cardiac cycle duration linearly correlates with the ECG, suggesting that the amount of PT delay is independent of the heart rate. This finding is consistent with the patient-specific bivariate tiled histograms of PT delay and ECG cardiac cycle duration in Supporting Information Fig. S2, which also shows that there is no significant dependence of PT delay on the ECG RR interval. These results are as expected since the cardiac PT triggers were identified as the negative peak of the first-order finite-difference, which refer to the phase of rapid mechanical activity during systole. The systolic part of the cardiac cycle only scales weakly with the cycle duration for a specific patient.

Fig. 7 shows examples of false ECG and PT triggers from two patients. For patient #1, Fig. 7(a) shows one false positive ECG trigger and two false positive PT triggers potentially due to the involuntary patient motion. This is not surprising because the performance of blind source separation methods can deteriorate with the inclusion of additional sources. Fig. 7(b) demonstrates the result of one slice from patient #21, in which ECG shows three false positive triggers. For this patient, severe arrhythmia can be observed in the cine images. Despite the weak and irregular heart rhythm, the intermediate PT signal strongly correlates with the cardiac motion seen in the temporal profiles.

## Discussion

PT is an emerging technology that allows the collection of physiological information during the MRI acquisition without pulse sequence modification or disruption of magnetization steady-state. With the PT generator integrated within the body array coil, this technique does not require additional patient preparation time. In this study, we have evaluated the accuracy and reliability of both cardiac and respiratory signals extracted from the PT data in patients clinically referred for CMR. While other studies have demonstrated successful extraction of respiratory and cardiac information from the PT signal, this is the first study that provides a direct comparison with both ECG and BM and uses information from RT cine images as the ground truth. Also, different from the previous studies, the potential for PT to be used as an accurate cardiac and respiratory gating signal is demonstrated in a patient cohort that includes subjects with irregular respiratory patterns and arrhythmia.

We have demonstrated that the respiratory motion can be extracted from the PT signal reliably using a combination of low-pass filtering and PCA. We find that the performance of PT is more robust compared to that of BM. PT shows a strong correlation with the reference for all cases except six slices (out of 287). Among them, the poor correlation of the two slices is due to the absence of respiratory motion, where the extracted respiratory signal captured the residual noise. In contrast, BM exhibits poor correlation in a number of slices due to the relative delay, incorrect directionality, or overall poor quality of the respiratory signal from BM. The consistency of the PT-driven respiratory binning with that of the reference is also calculated across all the patients, and our results are consistent with the findings from a recent study on 5D flow imaging [14].

To extract the cardiac motion, we employed both PCA and fast ICA [15]. The overall performance of PT-based cardiac trigger extraction is comparable to that of ECG. Among the six patients with arrhythmias (#2, #6, #10, #21, #22, and #23), PT had fewer false positive and negative triggers except for patient #22. Involuntary motion was observed for one patient (#1); here, ECG was found to be more robust than PT. Using the proposed pipeline, the cardiac information was extracted automatically for all the patients except for two cases (#21 and #22). Both of these patients had visibly weaker cardiac contractions (systolic dysfunction) along with arrhythmias. For these two subjects, we adjusted the frequency band of the filter and the threshold of negative peaks in the trigger detection process.

In this study, the cardiac PT triggers were identified as the negative peak of the first-order finite-difference. Therefore, they refer to the phase of rapid mechanical activity during systole. In contrast, ECG R-wave triggers correspond to the beginning of systole. Therefore, it is not surprising that PT triggers exhibit an average positive delay relative to the corresponding ECG triggers. The amount of delay has a dependence on the length of the systolic phase and thus varies from subject to subject, which partially accounts for the variations in PT delay in Table. 2. However, for a specific subject, the PT delay is independent of the heart rate since the systolic part of the cardiac cycle only scales weakly with the cycle duration. For PT-based retrospective triggering or post-acquisition binning of the CMR data, this delay should pose no issues other than changing the starting point of the cardiac cycle. For PT-guided prospective triggering, the impact of this delay needs to be considered, depending on the application [24]. Also, by anchoring the PT trigger to a feature other than the negative peak of the first-order finite-difference, it might be possible to minimize this delay. Compared to ECG, PT trigger jitter is 6.6–83.3 ms (median 21.8 ms). Not surprisingly, the worst performance is observed in patients with serious arrhythmia or weak cardiac contraction. We also investigated the dependence of PT delay and PT jitter on the patients' BMI/cardiac cycle duration, and no significant ($p > 0.28$) linear relationship is observed in this study.

We recognize some limitations of this study. First, all data processing in this work was performed retrospectively. Although preliminary studies suggest that PT can be used for prospective triggering [16, 17], additional investigations are required to establish prospective triggering capabilities of PT in a clinical scenario. Second, parameter tuning was needed to extract the cardiac motion reliably for the subjects with severe arrhythmia and very weak heart contractions. Further efforts are required to automate this process fully. Third, with our current implementation, the PT samples are available only when the sequence is acquiring MR data. The continuous sampling of the PT signal is feasible and will be considered in future studies. Our future work will focus on the application of the

PT-guided acquisition and reconstruction for various CMR applications, such as PT-guided 3D cardiac cine and 4D flow imaging.

## Conclusion

This study demonstrates the potential of PT to monitor both the respiratory and cardiac motion in patients clinically referred for CMR. The respiratory motion extracted from PT shows a strong agreement with image-derived respiratory motion. For the cardiac signal, PT is comparable with ECG in terms of trigger detection.

**Ethics approval and consent to participate**

The data were collected at xxxx with the ethical approval for recruitment and consent given by an Internal Review Board (xxxx)

**Availability of data and materials**

The datasets used and/or analyzed during the current study are available from the corresponding author upon requests

Table 1: Basic patient demographics. Fourteen male and nine female patients were included in this study, with an age range of 20-75 years. The patients were numbered based on the order of the acquisition. The body surface area (BSA), body mass index (BMI) and the corresponding CMR indications are shown. HCM: hypertrophic cardiomyopathy; PVC: premature ventricular contraction.

| Patient | Gender | Age [yr] | BSA [m$^2$] | BMI [Kg/m$^2$] | CMR Indications |
|---|---|---|---|---|---|
| 1 | Male | 52 | 2.09 | 30.27 | Ischemia evaluation |
| 2 | Female | 47 | 1.85 | 28.32 | Pulmonary Hypertension |
| 3 | Male | 58 | 2.39 | 29.21 | Atrial Fibrillation |
| 4 | Female | 22 | 1.59 | 23.78 | Right Ventricular Dilation |
| 5 | Male | 44 | 2.59 | 35.05 | Cardiomyopathy |
| 6 | Male | 75 | 2.47 | 34.79 | Cardiomyopathy, PVCs |
| 7 | Female | 52 | 1.78 | 27.39 | Ischemia evaluation |
| 8 | Female | 40 | 1.70 | 29.29 | Mitral Regurgitation |
| 9 | Male | 47 | 2.04 | 33.89 | HCM |
| 10 | Female | 74 | 2.04 | 37.76 | Ischemia evaluation |
| 11 | Female | 71 | 1.58 | 28.27 | Cardiomyopathy |
| 12 | Female | 41 | 2.28 | 39.53 | Intracardiac mass |
| 13 | Male | 53 | 1.80 | 21.41 | Cardiomyopathy |
| 14 | Female | 39 | 1.84 | 25.06 | PVCs |
| 15 | Male | 33 | 2.20 | 27.70 | PVCs |
| 16 | Male | 62 | 2.32 | 33.23 | Cardiomyopathy |
| 17 | Male | 62 | 1.77 | 28.70 | Cardiomyopathy |
| 18 | Male | 42 | 2.03 | 29.80 | HCM |
| 19 | Female | 59 | 1.69 | 24.37 | Cardiomyopathy |
| 20 | Male | 20 | 2.20 | 25.12 | Cardiomyopathy |
| 21 | Female | 49 | 1.75 | 24.96 | HCM |
| 22 | Male | 34 | 2.64 | 40.44 | Ischemia evaluation |
| 23 | Male | 61 | 1.76 | 21.29 | Cardiomyopathy |

Table 2: The consistency of the respiratory binning using the PT signal and the image-derived reference signal across all the patients. The k-space data were binned into four respiratory phases: end-expiratory, mid-expiratory, mid-inspiratory, and end-inspiratory. The average percentage of overlapping data for each pair of bins is reported here.

|  |  | Reference respiratory binning | | | |
|---|---|---|---|---|---|
|  | % | **End-exp** | **Mid-exp** | **Mid-insp** | **End-insp** |
| **PT respiratory binning** | **End-exp** | **80.03** | 17.95 | 1.92 | 0.1 |
|  | **Mid-exp** | 18.2 | **69.05** | 12.1 | 0.64 |
|  | **Mid-insp** | 1.57 | 12.43 | **75.79** | 10.21 |
|  | **End-insp** | 0.2 | 0.57 | 10.19 | **88.85** |

Table 3: The number of ECG triggers, the mean ECG RR interval, the number of false positive (FP) and false negative (FN) for ECG and PT triggers, the number of indeterminate triggers, and mean ± standard deviation (std) of PT delay. For each patient, data from all slices were aggregated.

| Patient | Number of ECG triggers | Mean ECG RR interval [ms] | FN ECG | FP ECG | FN PT | FP PT | Indeter-minate | PT delay [ms] |
|---|---|---|---|---|---|---|---|---|
| 1 | 120 | 1073.5 | 0 | 1 | 0 | 2 | 0 | 166.2 ± 38.6 |
| 2 | 145 | 847.7 | 0 | 0 | 0 | 0 | 0 | 166.5 ± 11.2 |
| 3 | 153 | 1161.9 | 0 | 0 | 0 | 0 | 0 | 195.2 ± 16.8 |
| 4 | 105 | 695.1 | 0 | 0 | 0 | 0 | 0 | 120.9 ± 31.0 |
| 5 | 156 | 809.2 | 0 | 0 | 0 | 0 | 0 | 174.8 ± 18.2 |
| 6 | 110 | 1140.2 | 2 | 0 | 0 | 0 | 2 | 146.4 ± 19.6 |
| 7 | 137 | 820.5 | 0 | 1 | 0 | 0 | 0 | 206.2 ± 18.3 |
| 8 | 165 | 691.7 | 0 | 0 | 0 | 0 | 1 | 113.2 ± 25.2 |
| 9 | 145 | 929.5 | 0 | 0 | 4 | 1 | 0 | 79.3 ± 39.2 |
| 10 | 151 | 901.6 | 4 | 1 | 0 | 0 | 1 | 133.1 ± 21.4 |
| 11 | 143 | 822.9 | 0 | 0 | 0 | 0 | 0 | 206.2 ± 6.6 |
| 12 | 92 | 787.4 | 0 | 0 | 0 | 0 | 0 | 158.5 ± 22.4 |
| 13 | 122 | 937 | 0 | 1 | 0 | 0 | 0 | 142.0 ± 19.6 |
| 14 | 143 | 803.8 | 0 | 0 | 0 | 0 | 0 | 143.1 ± 21.1 |
| 15 | 116 | 1205.8 | 0 | 0 | 0 | 0 | 0 | 144.8 ± 21.8 |
| 16 | 121 | 941.5 | 0 | 0 | 0 | 0 | 0 | 176.7 ± 23.8 |
| 17 | 155 | 721.1 | 0 | 1 | 0 | 0 | 0 | 107.4 ± 12.5 |
| 18 | 145 | 779.5 | 0 | 0 | 1 | 1 | 0 | 116.3 ± 32.2 |
| 19 | 154 | 749.1 | 0 | 0 | 0 | 0 | 0 | 164.7 ± 29.7 |
| 20 | 159 | 855 | 0 | 0 | 0 | 0 | 0 | 125.1 ± 20.6 |
| 21 | 208 | 591.7 | 0 | 10 | 6 | 1 | 6 | 179.7 ± 83.3 |
| 22 | 209 | 628.4 | 0 | 2 | 2 | 3 | 1 | 143.4 ± 47.6 |
| 23 | 153 | 809.2 | 14 | 0 | 0 | 1 | 1 | 220.1 ± 28.8 |
| Total | 3307 | — | 20 | 17 | 13 | 9 | 11 | — |

Figure 1: Pipeline to extract respiratory and cardiac motions from the multi-coil PT signal with C coils and N samples for each coil. The respiratory motion was extracted using a low-pass filter and PCA. From the first two principal components, the component with the highest peak between [0.1, 0.6] Hz was selected. The expiration (E) and inspiration (I) were automatically identified based on the polarity of the PT signal from the spine coils closest to the feet. To extract the cardiac motion, both PCA and fast ICA were utilized, and the independent component with the highest peak in the [0.5, 3] Hz frequency band was selected as the cardiac signal. The resulting cardiac signal was then further band-pass filtered to suppress noise. After sign correction, the trigger points were detected as the negative peak in the first-order finite-difference of the extracted signal.

Figure 2: Pearson correlation coefficients of the PT and BM signals with the image-derived reference signal. As highlighted by the red arrows, there are six slices (out of 287) where PT poorly correlated (absolute correlation coefficient less than 0.7) with the reference. In contrast, 81 slices for BM exhibited absolute correlation coefficients of less than 0.7.

Figure 3: A direct comparison of the respiratory signals extracted from PT (blue), BM (black), and the image-derived reference (red) from six patients. The temporal profiles along the green dashed lines drawn through the liver dome are shown for visual reference. The PT signal strongly correlated with the reference for all the patients in the top two rows. However, the performance of BM was unstable. There is no respiratory motion for patient #21; therefore, both PT and BM poorly correlated with the reference even though they captured the lack of respiratory motion correctly. For patient #18, a significant mismatch with the reference was observed for both PT and BM.

Figure 4: A direct comparison between cardiac PT (orange) and ECG (black) traces for a represented patient. The temporal profile along the green dashed line is shown for visual reference, with the end diastolic frames marked with green triangles. The noisy blue line in the middle row is the intermediate PT signal after PCA and fast ICA, which exhibits the modulation due to the cardiac pulsation. The final PT signal (orange) is obtained by first filtering and then taking the first-order finite-difference of the signal in blue. The PT triggers (red stars) are identified as the negative peaks.

Figure 5: Pairing process of the PT and ECG triggers. (a) The PT triggers were paired with the corresponding ECG triggers. (b) All the unpaired triggers were sorted into five groups based on the number of systoles recognized by a cardiologist from the corresponding cine images: false positive/negative ECG, false positive/negative PT, or indeterminate. The top row in (b) is the RT image or temporal profiles along the line drawn through the middle of the heart.

Figure 6: Direct comparison of the cardiac interval derived from the PT and ECG triggers. Overall, the PT cardiac interval is in good agreement with the ECG, with the mean absolute difference less than 5% of the averaged ECG cardiac interval, except for patients #21 (14%) and #22 (7.7%).

Figure 7: Examples of false ECG and PT triggers from two patients. (a) False positive PT/ECG triggers from patient #1. There are two false positive PT (red triangles) and one false positive ECG (purple triangle) triggers due to another source of significant motion, probably a hiccup (yellow arrow). (b) False positive ECG triggers from patient #21. There are three false positive ECG triggers highlighted by the purple triangles. Due to the severe arrhythmia, the frequency band of the filter and the threshold of negative peaks in the trigger detection process were adjusted to extract the cardiac modulation and identify the cardiac PT triggers successfully for patient #21.

Supporting Information Figure S1: The dependence of PT delay and PT jitter on the patients' BMI/cardiac cycle duration, where patient #21 (red star) is excluded due to high arrhythmic load and weak cardiac contraction. No significant linear relationship is observed since the correlation coefficient is not significantly ($p > 0.28$) different from zero. The red dashed curve is the confidence bounds with 95% confidence.

Supporting Information Figure S2: The bivariate tiled histograms of PT delay and ECG cardiac cycle duration. The bin colors represent the number of data points inside the bin with darker blue indicating smaller values. Note, the color map is scaled differently across patients to improve visualization. The mean and standard deviation of PT delay are shown in the top left corner. For a specific patient, there is no significant dependence of PT delay on the ECG RR interval.

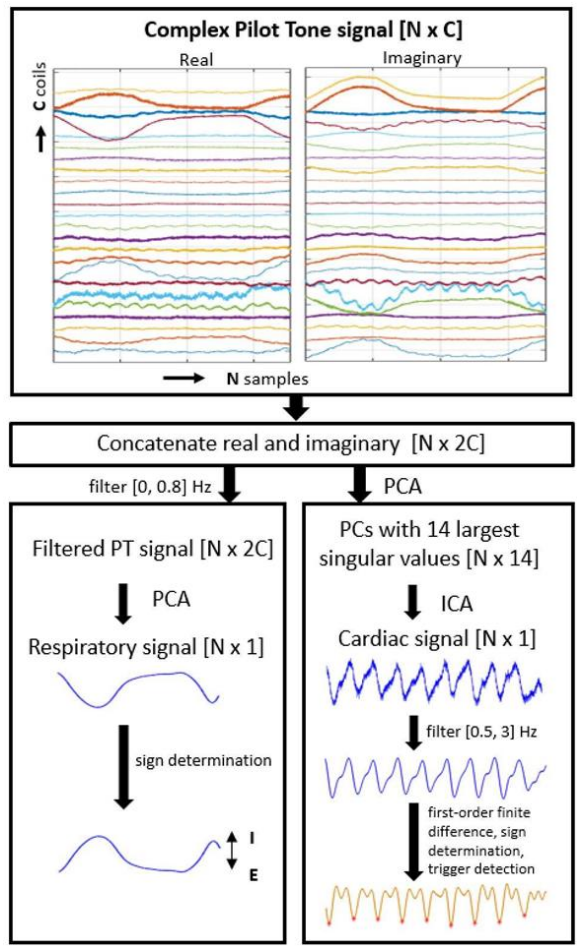

Figure 1 Pipeline to extract respiratory and cardiac motions from the multi-coil PT signal with C coils and N samples for each coil. The respiratory motion was extracted using a low-pass filter and PCA. From the first two principal components, the component with the highest peak between [0.1, 0.6] Hz was selected. The expiration (E) and inspiration (I) were automatically identified based on the polarity of the PT signal from the spine coils closest to the feet. To extract the cardiac motion, both PCA and fast ICA were utilized, and the independent component with the highest peak in the [0.5, 3] Hz frequency band was selected as the cardiac signal. The resulting cardiac signal was then further band-pass filtered to suppress noise. After sign correction, the trigger points were detected as the negative peak in the first-order finite-difference of the extracted signal.

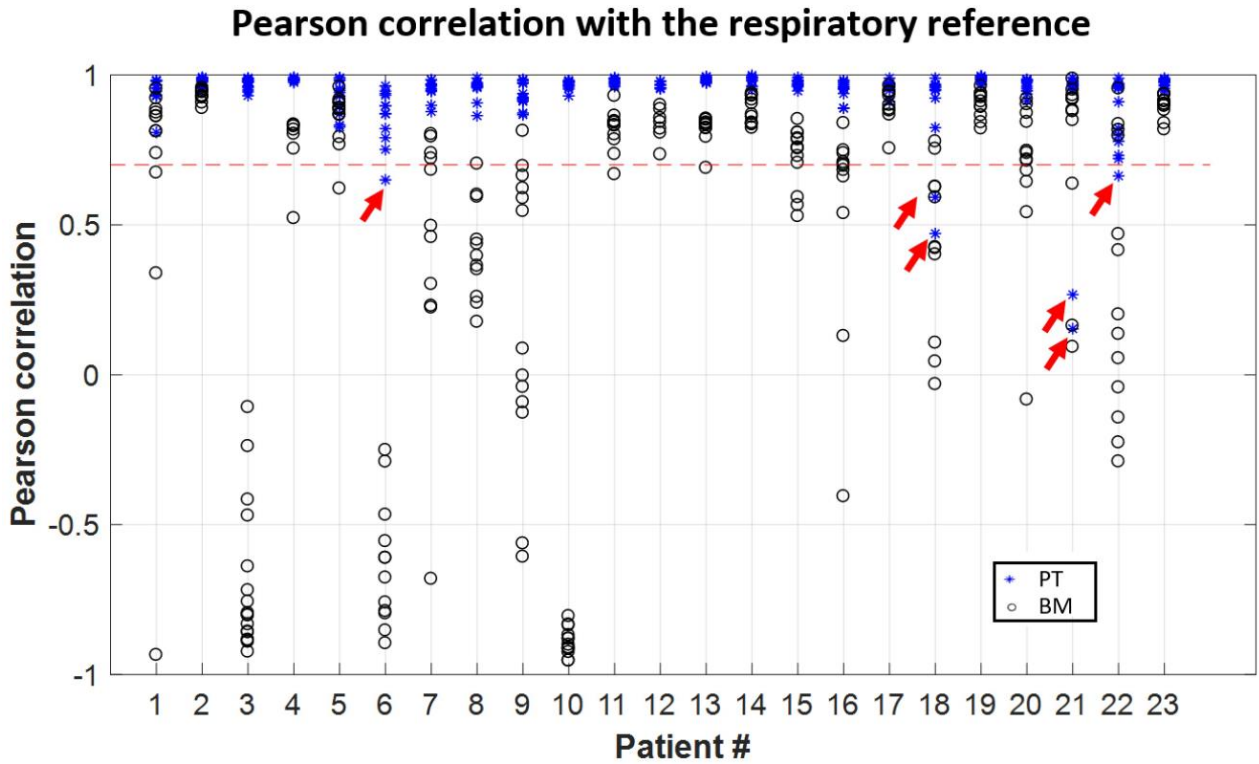

Figure 2: Pearson correlation coefficients of the PT and BM signals with the image-derived reference signal. As highlighted by the red arrows, there are six slices (out of 287) where PT poorly correlated (absolute correlation coefficient less than 0.7) with the reference. In contrast, 81 slices for BM exhibited absolute correlation coefficients of less than 0.7.

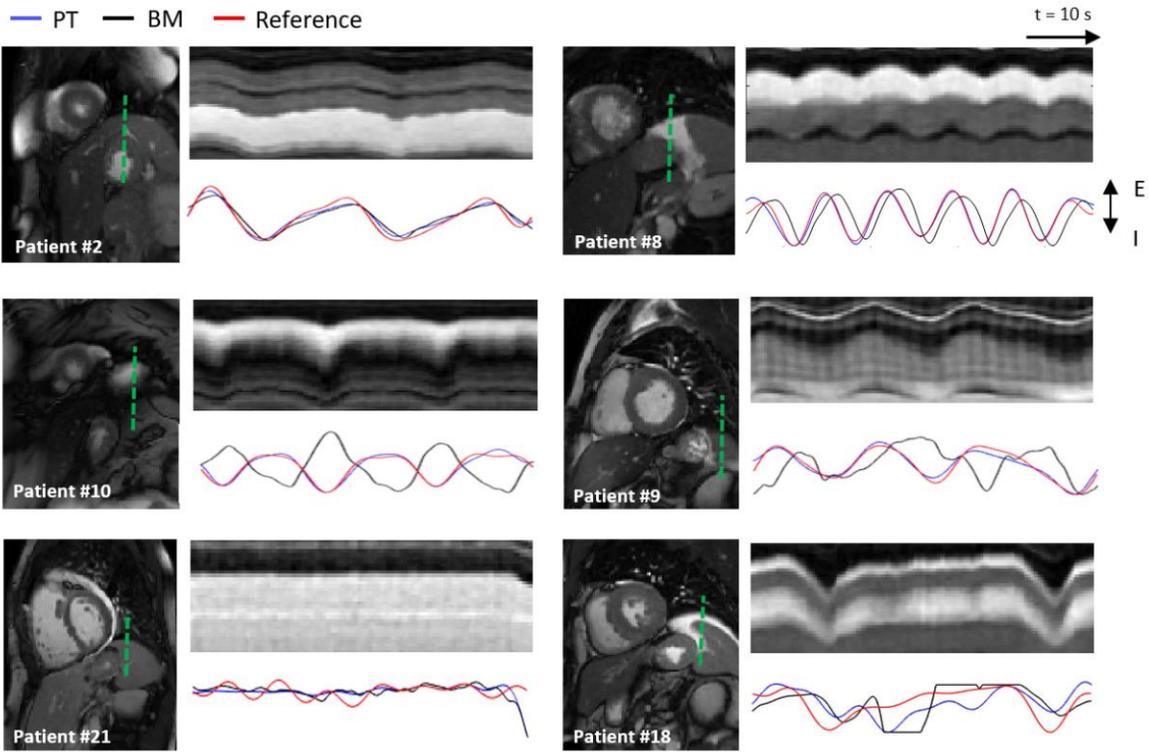

Figure 3: A direct comparison of the respiratory signals extracted from PT (blue), BM (black), and the image-derived reference (red) from six patients. The temporal profiles along the green dashed lines drawn through the liver dome are shown for visual reference. The PT signal strongly correlated with the reference for all the patients in the top two rows. However, the performance of BM was unstable. There is no respiratory motion for patient #21; therefore, both PT and BM poorly correlated with the reference even though they captured the lack of respiratory motion correctly. For patient #18, a significant mismatch with the reference was observed for both PT and BM.

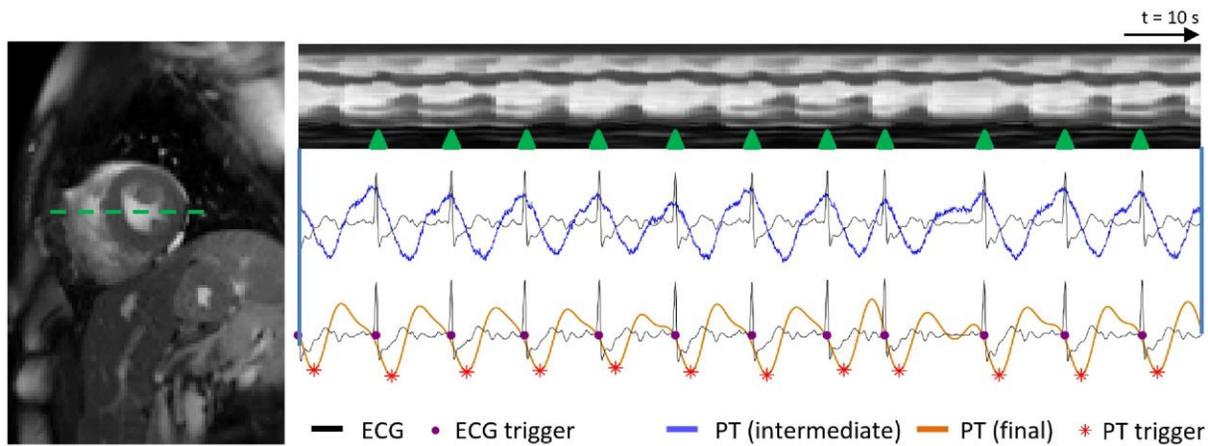

Figure 4: A direct comparison between cardiac PT (orange) and ECG (black) traces for a represented patient. The temporal profile along the green dashed line is shown for visual reference, with the end-diastolic frames marked with green triangles. The noisy blue line in the middle row is the intermediate PT signal after PCA and fast ICA, which exhibits the modulation due to the cardiac pulsation. The final PT signal (orange) is obtained by first filtering and then taking the first-order finite-difference of the signal in blue. The PT triggers (red stars) are identified as the negative peaks

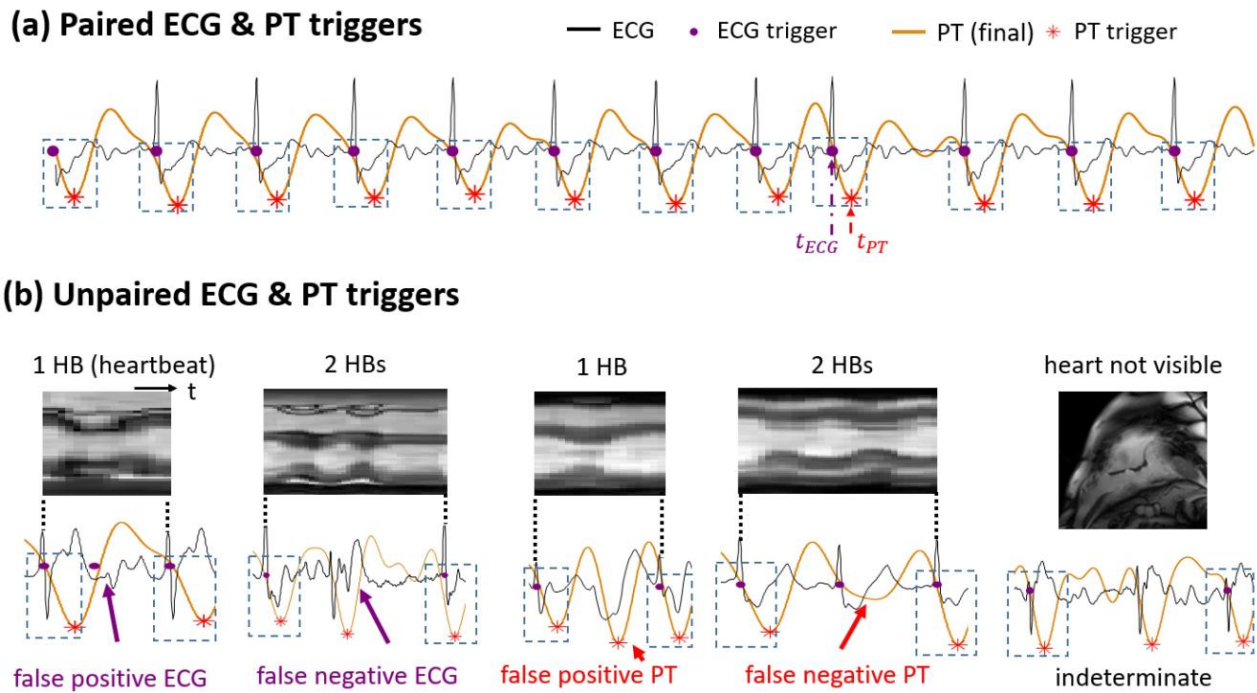

Figure 5: Pairing process of the PT and ECG triggers. (a) The PT triggers were paired with the corresponding ECG triggers. (b) All the unpaired triggers were sorted into five groups based on the number of systoles recognized by a cardiologist from the corresponding cine images: false positive/negative ECG, false positive/negative PT, or indeterminate. The top row in (b) is the RT image or temporal profiles along the line drawn through the middle of the heart.

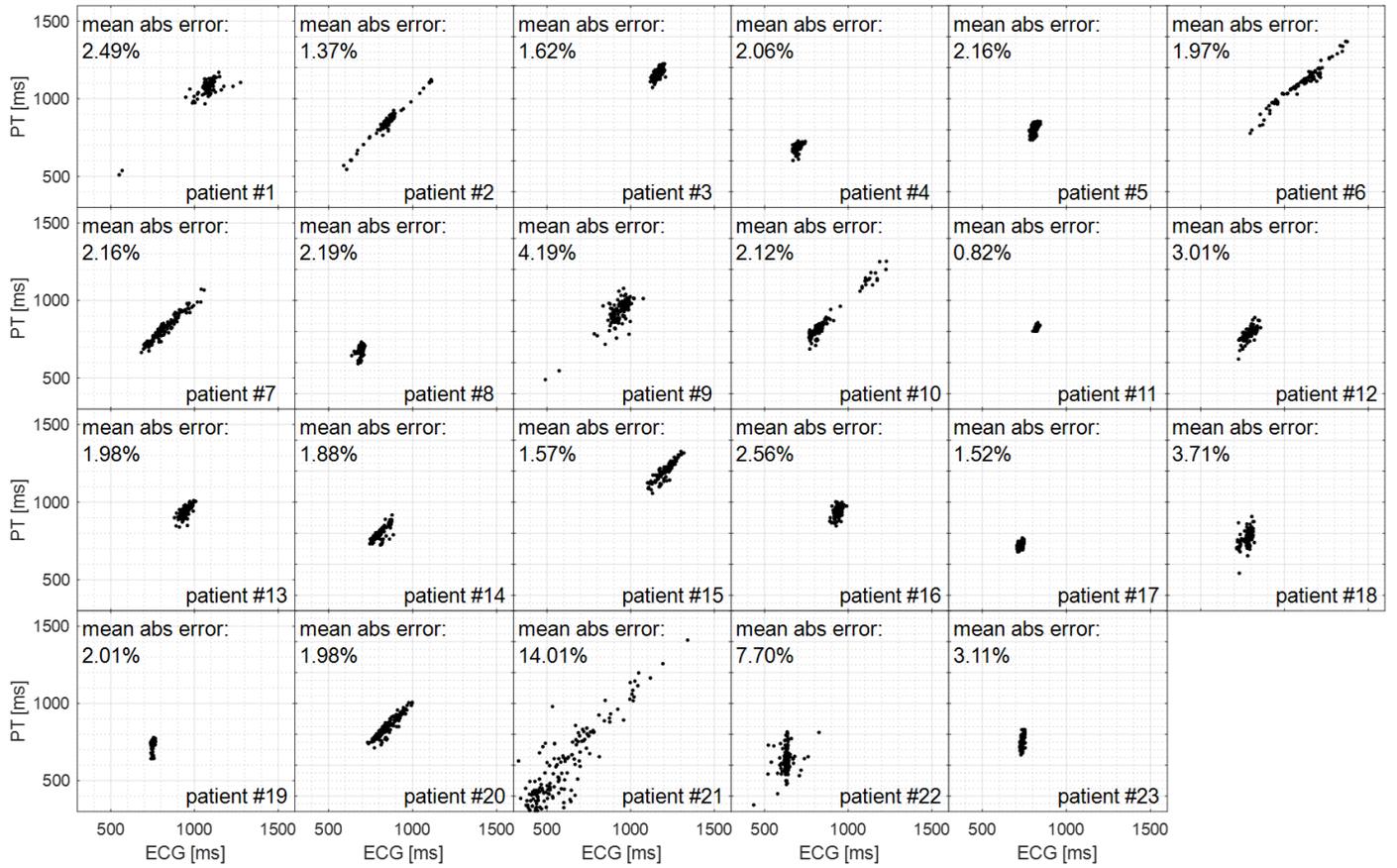

Figure 6: Direct comparison of the cardiac cycle duration estimated from the PT and corresponding ECG triggers. Overall, the PT cardiac cycle duration is in good agreement with the ECG, with the mean absolute difference less than 5% of the averaged ECG RR interval, except for patients #21 (14%) and #22 (7.7%).

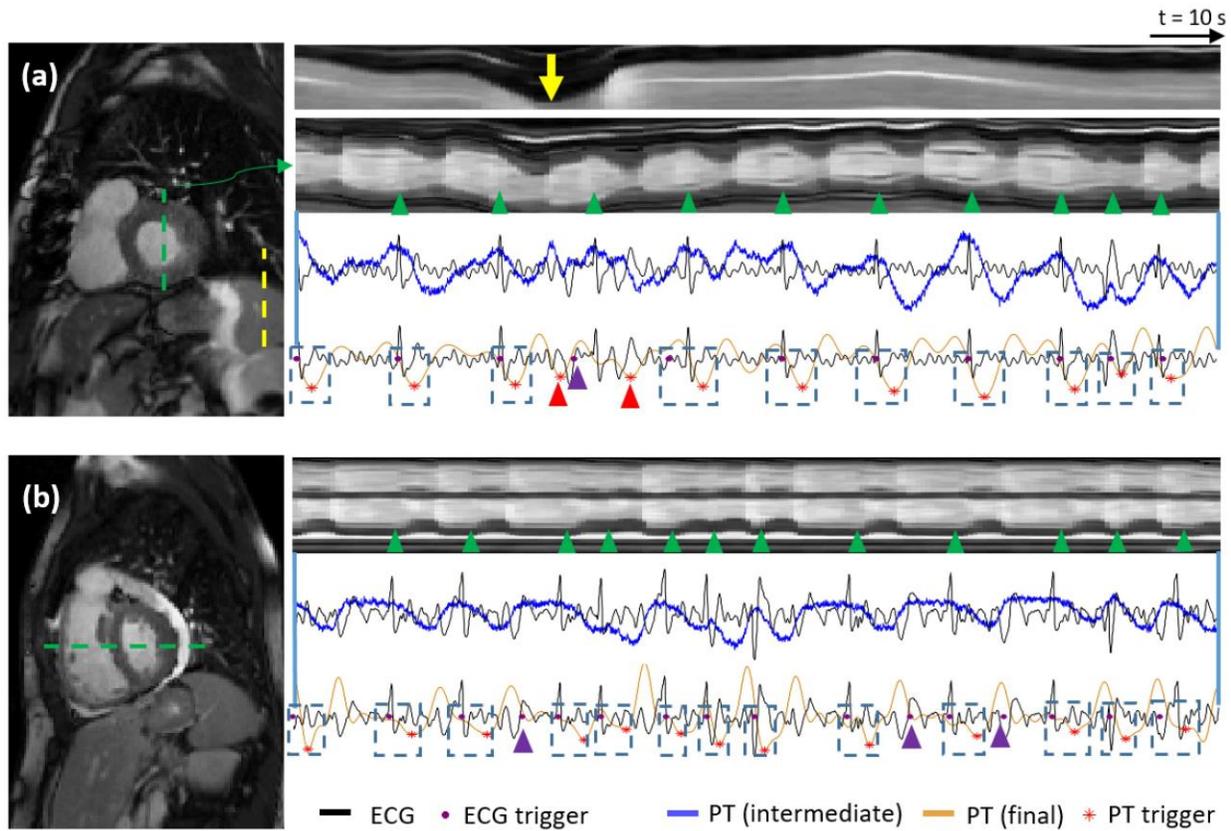

Figure 7: Examples of false ECG and PT triggers from two patients. (a) False positive PT/ECG triggers from patient #1. There are two false positive PT (red triangles) and one false positive ECG (purple triangle) triggers due to another source of significant motion, probably a hiccup (yellow arrow). (b) False positive ECG triggers from patient #21. There are three false positive ECG triggers highlighted by the purple triangles. Due to the severe arrhythmia, the frequency band of the filter and the threshold of negative peaks in the trigger detection process were adjusted to extract the cardiac modulation and identify the cardiac PT triggers successfully for patient #21.

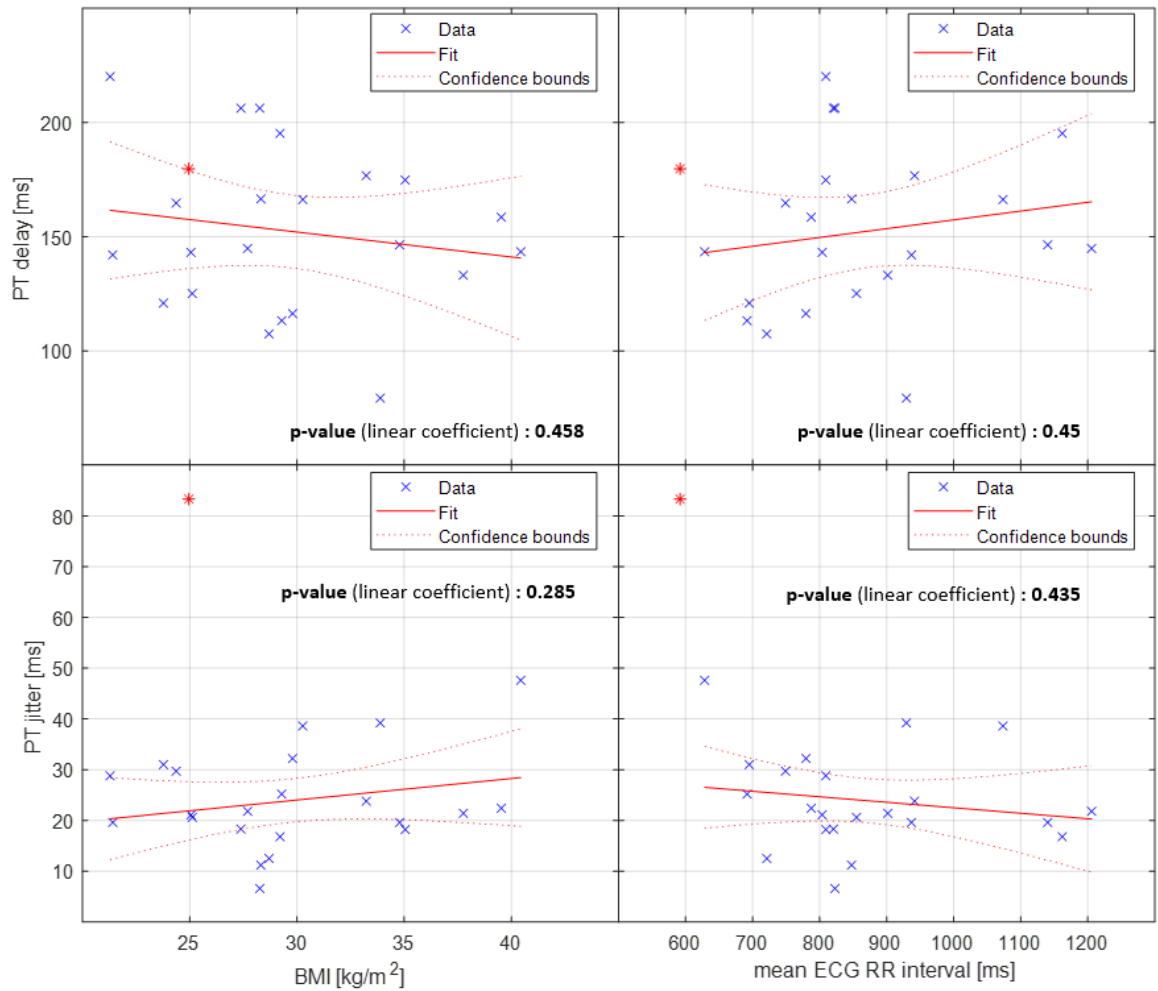

Supporting Information Figure S1: The dependence of PT delay and PT jitter on the patients' BMI/cardiac cycle duration, where patient #21 (red star) is excluded due to high arrhythmic load and weak cardiac contraction. No significant linear relationship is observed since the correlation coefficient is not significantly (p > 0.28) different from zero. The red dashed curve is the confidence bounds with 95% confidence.

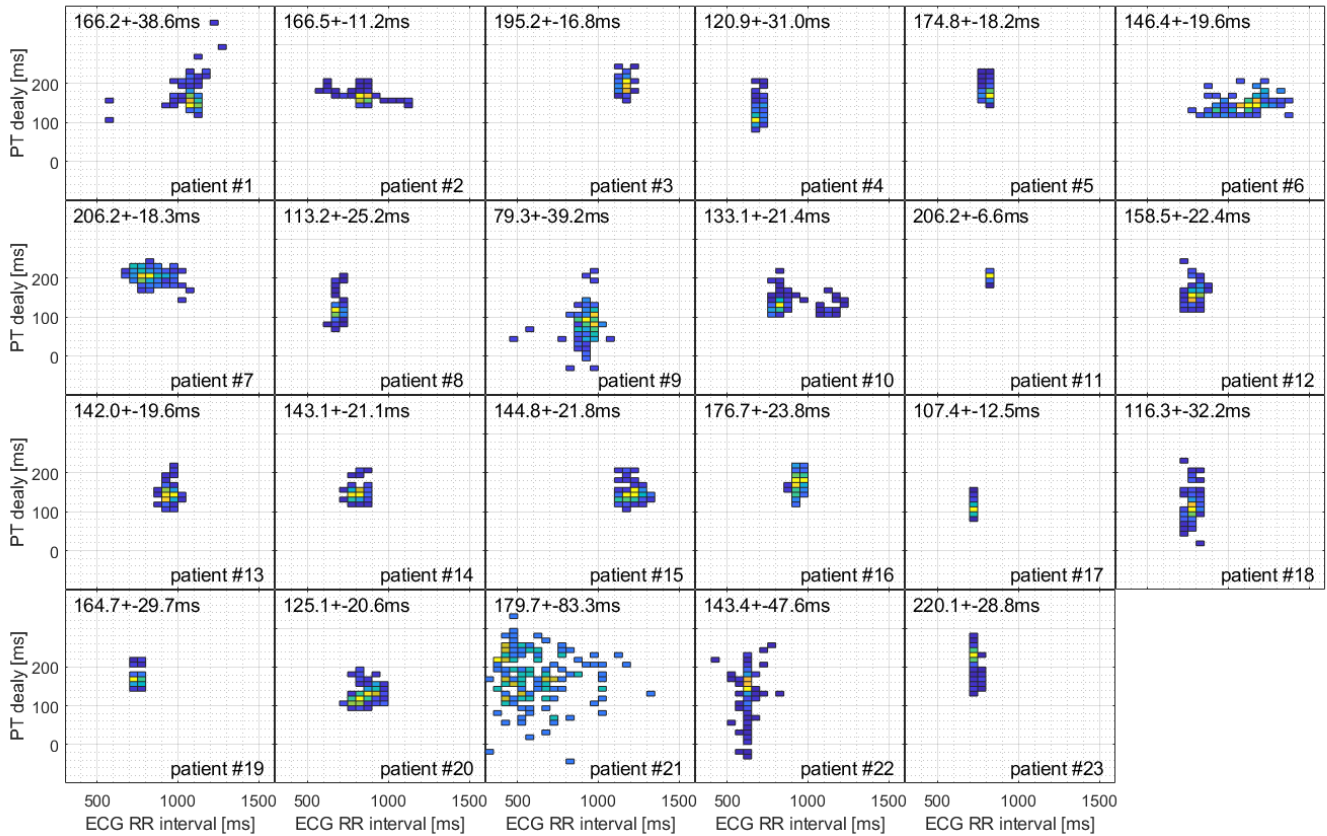

Supporting Information Figure S2: The bivariate tiled histograms of PT delay and ECG cardiac cycle duration. The bin colors represent the number of data points inside the bin with darker blue indicating smaller values. Note the color map is scaled differently across patients to improve visualization. The mean and standard deviation of PT delay are shown in the top left corner. For a specific patient, there is no significant dependence of PT delay on the ECG RR interval.